          [1999/12/02 1.01 (PWD)]
\documentclass[letterpaper,twocolumn]{esapub} 

\usepackage{natbib}
\usepackage{epsfig}
\usepackage{graphics}

\title{Radio Emission from the X-ray transient XTE J1550$-$564}
\author[1]{Diana Hannikainen}
\affil{Department of Physics and Astronomy, University of Southampton,
SO17 1BJ Southampton, UK}
\author[2,3]{Kinwah Wu}
\affil[2]{School of Physics, University of Sydney, NSW 2006, Australia}
\affil[3]{MSSL, University College London, Holmbury St. Mary,
Dorking, Surrey RH5 6NT, UK}

\author[2]{Duncan Campbell-Wilson}
\author[2]{Richard Hunstead}  
\author[4]{\\ Jim Lovell}
\author[    \,2]{Vince McIntyre\footnote{Present address:
ATNF, CSIRO, PO Box 76, Epping, NSW 1710, Australia}}
\author[4]{John Reynolds}
\author[3]{Roberto Soria}
\author[4]{Tasso Tzioumis}
\affil[4]{ATNF, CSIRO, PO Box 76, Epping, NSW 1710, Australia}

\newcommand{\xtej}{XTE\,J1550$-$564~}

\begin{document}


\maketitle

\begin{abstract}

We report multifrequency radio observations of \xtej obtained with
the Molonglo Observatory Synthesis Telescope and the Australia
Telescope Compact Array at the time of its discovery and subsequent
hard and soft X-ray outburst in 1998 September. A large radio flare
was observed, peaking about 1.8 days after the X-ray
flare. In addition, we present Australian Long Baseline Array 
images obtained shortly after the maximum of the radio flare which 
show evolving structure. The apparent separation velocity of
the two outermost components is $v>2c$.

\end{abstract}

\section{Introduction}

The soft X-ray transient \xtej was discovered by the 
  All-Sky Monitor (ASM) on board 
  the Rossi X-ray Timing Explorer (RXTE) on MJD~51063 (1998 
  Sept 7; MJD=JD$-$240000.5) with 
  an intensity of $\sim$0.07~Crab in the 2--12~keV range \cite{smith98}
  and by the Burst and Transient Source Experiment
  (BATSE) on the Compton Gamma-Ray Observatory
   in the 20--100~keV range at a flux level of ${1.26 \times 10^{-8} \
  \rm erg} \ {\rm
  s}^{-1} \ {\rm cm}^{-2}$ \cite{wilson98}. 
The RXTE/ASM intensity increased steadily over the next few days, 
  reaching $\sim$1.7~Crab on MJD~51071 \cite{rem98}.
\xtej flared to 6.8~Crab on MJD~51075--51076, making it the 
  brightest X-ray nova observed with RXTE to date \cite{rem98}.
An optical counterpart with a magnitude of V=16 was identified on 
  MJD~51065 (Orosz, Bailyn \& Jain 1998), and an orbital period of $\sim$1.54 days was
  attributed to the system \cite{jain}.
A radio counterpart was discovered at the optical position with a 
  flux density of 10$\pm$2.5~mJy on MJD~51065 \cite{cw98}. 
Evolving QPOs \cite{cui} and hard lags \cite{wijn} typical of
  black hole binaries were observed from XTE\,J1550$-$564.

In this paper we introduce briefly the radio emission from 
  \xtej at the time of the 1998 outburst. A more
  detailed analysis will be presented in Hannikainen et al. (2001, 
  in preparation).

\section{observations}

\begin{figure*}
\begin{center}
\epsfig{figure=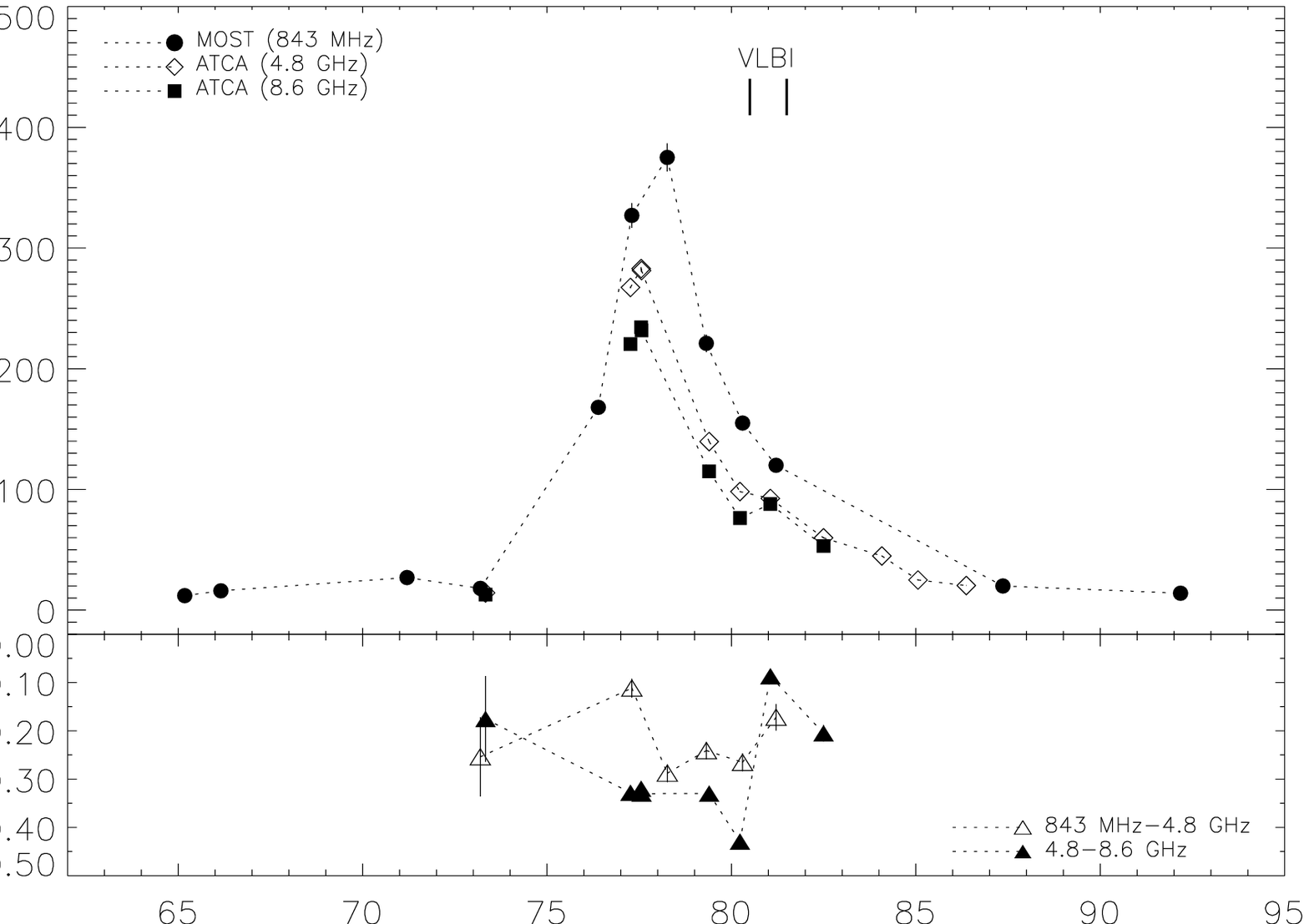,width=13cm}
\end{center}
\caption{The upper panel shows the MOST 843~MHz lightcurve (filled
circles) along with the ATCA 4.8~GHz (open diamonds) and 8.6~GHz
(filled squares) lightcurves.  The two vertical lines denote the
epochs of the VLBI observations. The bottom panel shows the 2-point
spectral indices derived from the 843~MHz--4.8~GHz data (open
triangles) and from the 4.8--8.6~GHz data (filled triangles).}

\begin{center}
\epsfig{figure=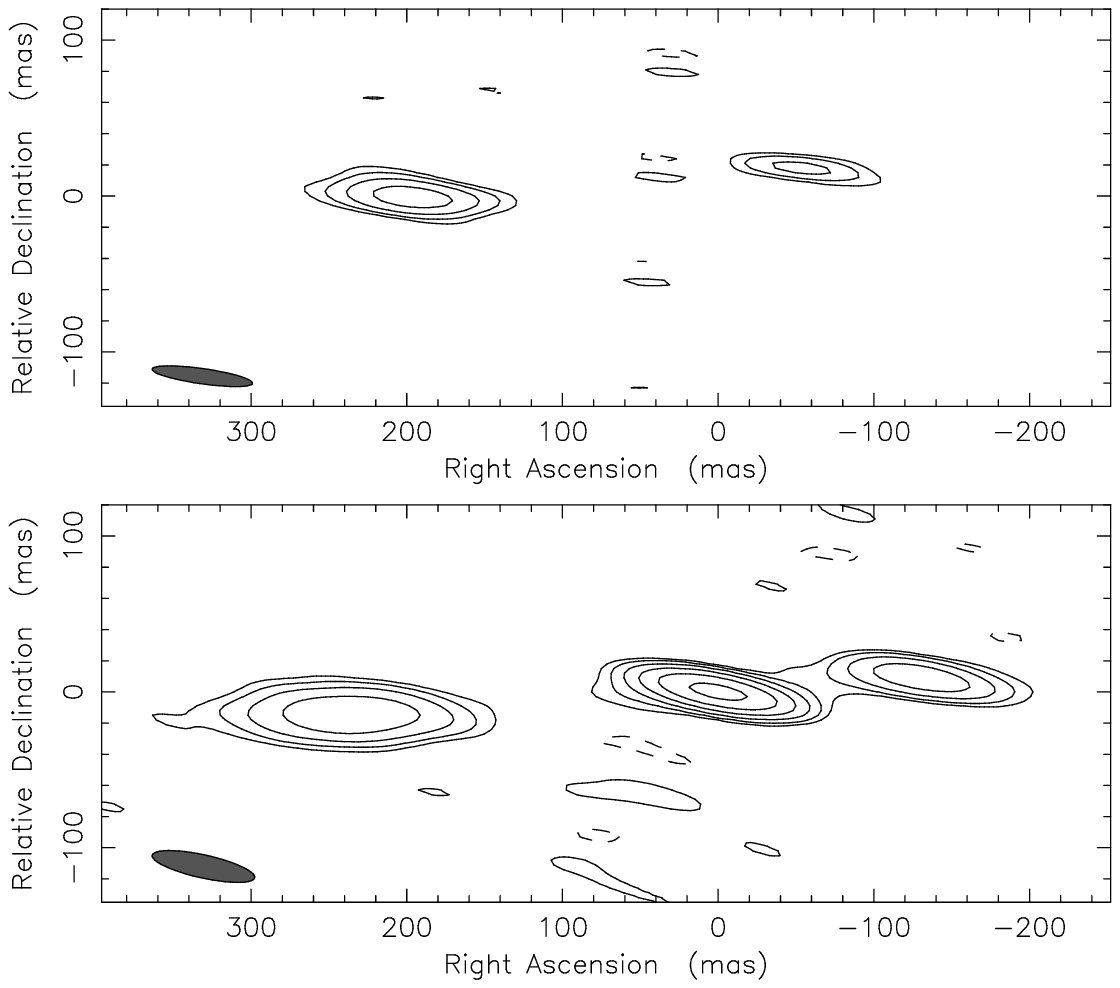, width=13cm}
\end{center}
\caption{The 2.29~GHz VLBI images from MJD~51080.5 (top) and
MJD~51081.5 (bottom).
The flux densities in the components, from east to west, 
are: 71~mJy and 20~mJy (top); 19~mJy, 25~mJy and 8~mJy
(bottom). The beam is shown in the lower left-hand corner.}
\label{fig2}
\end{figure*}

\begin{figure*}[t]
\begin{center}
\epsfxsize=13cm
\epsfbox{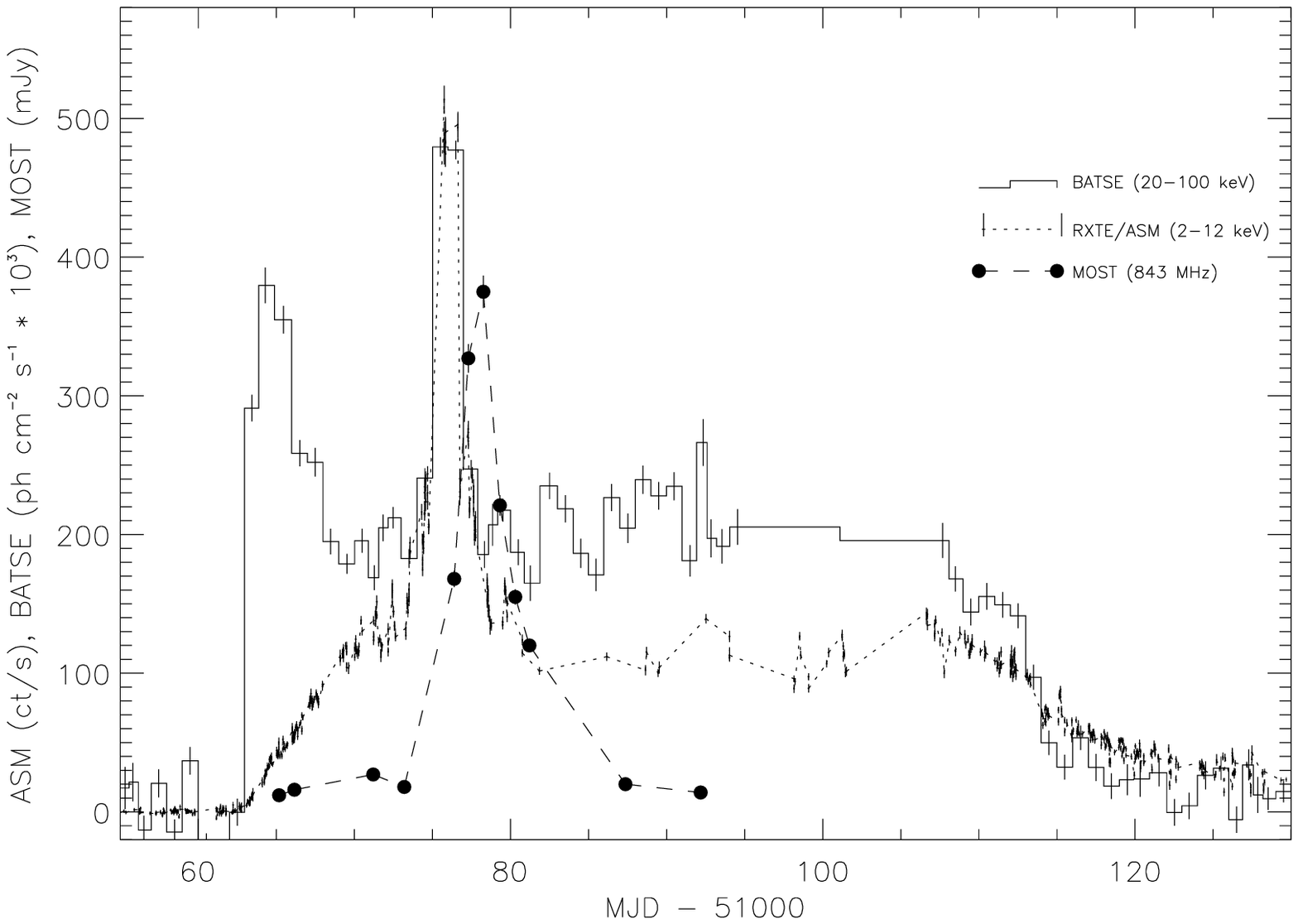}
\end{center}
\caption{The \xtej multiwavelength lightcurves: BATSE
20--100~keV (histogram),  RXTE/ASM 2--12~keV (
dotted line), MOST 843~MHz (filled circles and dashed line).  }
\label{fig1}
\end{figure*}

\xtej was observed at multiple radio frequencies during the 
  September 1998 outburst.
The Molonglo Observatory Synthesis Telescope (MOST)
  made twelve observations of the source
  at 843~MHz between MJD 51065 and 51092,
  while the Australia Telescope Compact Array (ATCA)
  made observations at 1.4, 2.3, 4.8 and 8.6~GHz between
  MJD~51073 and 51085.
In addition, Very Long Baseline Interferometry (VLBI) 
  images were obtained with the Australian Long Baseline Array (LBA)
  at 2.29~GHz.  
The MOST 843~MHz and ATCA 4.8 and 8.6~GHz lightcurves are shown in the 
  upper panel of Figure 1
  along with the epochs of the VLBI observations. 

\subsection{MOST \& ATCA}
Following the initial detection of \xtej on MJD~51065, the
  MOST monitored the source at 843~MHz over the next 27 days
  (Figure 1). 
Between MJD~51065 and 51073, the source flux density remained
  between 10 and 30~mJy. 
After MJD~51073 the flux density began to increase, reaching 
  168~mJy on MJD~51076, and peaking on MJD~51078 with 
  a flux density of 375~mJy. 
The flux density then declined to 120~mJy over the next
  three days, with a continuing decline to 
  14~mJy on MJD~51092, after which the monitoring ceased.

The ATCA started observing \xtej on MJD~51073 while the
  flux density was still low -- $10-12$~mJy at 4.8 and 8.6~GHz
  (Figure 1). 
After the rise to 168~mJy detected with the MOST, the ATCA resumed
  observing the source and monitored it over the next ten days.
The 4.8 and 8.6~GHz flux densities appear to peak approximately
  0.5~days before the MOST peak, with flux densities of 283 and 234~mJy
  respectively and linear polarization $\sim$~6\%.
The ATCA flux densities then followed the same decline as the MOST,
  reaching a level of around 20~mJy at 4.8~GHz on
  MJD~51086.5 when monitoring was ceased.
The linearly polarized flux also evolved during the outburst
  (Hannikainen et al., in preparation),
  reminiscent of the behavior of GRO~J1655$-$40 during the 1994 ejection 
  episodes \cite{hanni2000a}.

\subsection{VLBI}

The VLBI observations were undertaken on MJD~51080.5 and MJD~51081.5
  using the LBA \cite{hanni2000b}.
On the first day, the 34-m DSS45, 
  the 26-m Hobart and the 22-m Mopra antennas were
  used, while on the second day the 70-m DSS43 replaced the DSS45, and the
  six 22-m dishes of the ATCA were also used. 

Figure 2 shows the 2.29~GHz VLBI images obtained on MJD~51080.5 (top)
   and MJD~51081.5 (bottom).

The image from MJD~51081.5 shows three distinct components, 
  and the structure has 
  clearly evolved compared to the previous day.
Phase-referenced observations were not performed on either day so the
   zero coordinate in the images is arbitrary.
However, changes in the spectral index, 
  $\alpha$ ($S_{\nu}\propto\nu^{\alpha}$), are
  consistent with the appearance of a third component (Figure 1).
Although the overall 2-point spectral indices, both the
  843~MHz--4.8~GHz and the 4.8--8.6~GHz, stay relatively flat throughout
  the observing period (much flatter, say, compared to GRO\,J1655$-$40 which
  had $\alpha$ of $\sim \ -0.5$ to $-0.9$ during the 1994 ejection
  episodes; Hannikainen et al. 2000a) the 4.8--8.6~GHz spectral index does
  flatten from $-0.45$ to $-0.1$ just before the VLBI observations.
Thus, the central component in Figure 2 (bottom) is tentatively identified with
   a new optically thick outburst.
If this interpretation is correct,
   then the apparent separation velocity of the outermost
   components between MJD~51080.5 and 51081.5 is $v>2c$
   \cite{hanni2000b},
   based on a distance estimate of 3.5--5~kpc
   derived from H~{\sc i} observations.

\section{Multiwavelength Lightcurves}

Figure 3 shows the multiwavelength behavior of XTE\,J1550$-$564. 
The significance of the sequence of lightcurve features in each 
  waveband is discussed in Wu et al. (2000) and an in-depth 
  analysis of the multiwavelength data is in Hannikainen
  et al. (2001, in preparation), so here we shall 
  only summarize the more salient features. 
The outburst sequence is characterized by an impulsive rise in  
  hard X-rays accompanied by a gradual brightening in the soft 
  X-rays, followed by a period of exponential-like decay in the 
  hard X-ray luminosity, and then a giant flare in all wavebands. 
The hard and soft X-rays peaked simultaneously, while the radio flare
  maxima occurred about 1.3--1.8 days later.
As the radio emission continued to decline to a low level after the
  flare, the X-ray intensities settled into a plateau, lingered for
  another 35 days, and then faded away. 

In order to explain the first impulsive BATSE hard X-ray outburst,
  which does not conform to the standard disk-instability scenario, a
  model involving a magnetic secondary star is proposed \cite{wu}.
In this model, coronal magnetic activity lifts matter above the
  stellar surface of the secondary (which is presumed to be a G/K
  subgiant; see also Bildsten \& Rutledge 1999), and if this happens
  near the L$_1$ point of the Roche lobe, this matter will fall easily
  into the Roche lobe of the black hole, as gravity is weak near the
  L$_1$ point. 
Provided the magnetic stress/tension is stronger than the gas pressure
  terms, matter can be trapped there and accumulate over time. 
Eventually, this magnetic dam may burst leading to field-line breaking
  and a matter avalanche. 
The initial angular momentum of the matter with respect to the black
  hole is insignificant, as the matter is practically in free-fall
  from zero velocity. 
This flow will therefore be quasi-spherical, and deviation occurs
  only when approaching the black hole. 
An accretion shock forms when the infalling material encounters
  the centrifugal barrier of the hole, converting a portion of the
  kinetic energy to thermal energy, giving rise to radiation. 
These photons are then upscattered by the surging infalling matter to
  hard X-ray energies, which is observed as the first hard X-ray
  outburst. 
As the quasi-spherical flow subsides, the hard X-ray luminosity
  declines, and at the same time, residual matter with substantial
  angular momentum begins to condense in the equatorial plane to form
  an accretion disk -- this is seen as the gradual rise in the soft
  X-rays (see Igumenschev, Illarionov \& Abramowicz 1999). 
Irradiation of the secondary as a consequence of the first hard X-ray
  outburst will result in an increase of mass transfer, and provoke
  instabilities, seen in the large fluctuations in the soft X-ray
  luminosity between MJD~51070 and 51075.
As a result, the accretion disk becomes unstable and portions of it
  collapse, leading to mass ejection, as evidenced by the radio flare
  and the discrete components observed with the VLBI.

\section{Summary}

The \xtej hard and soft X-ray flare in 1998 September was accompanied 
by radio emission. The radio
outburst, which reached 375~mJy at 843~MHz, 
lagged the X-ray flare by 1.8 days. There was spectral
evolution during the outburst, and polarization was detected at the
6\% level. The VLBI images showed evolving structure confirming that
there was an outflow from the system. 
The inferred superluminal separation velocity of the two outermost
 components, $v > 2c$, strengthens the argument for XTE\,J1550$-$564
 being a black-hole binary system similar to GRO\,J1655$-$40 and GRS\,1915+105.

\section*{Acknowledgments}

We thank Michael McCollough and the BATSE team for the BATSE data, and 
the referee Chris Shrader for helpful suggestions.
DCH acknowledges the support of a PPARC postdoctoral research grant to
the University of Southampton, and financial support from the Academy
of Finland, and thanks the Astrophysics Department, University of
Sydney for hospitality during her visits. 
KW thanks Prof. P.~Charles for funding his visits to the University of
Southampton.
MOST is operated by the University of Sydney and supported by grants
from the Australian Research Council. The Australia Telescope Compact
Array is funded by the Commonwealth of Australia for operation as a
National Facility managed by CSIRO.
This research has made use of data obtained from the High Energy
Astrophysics Science Archive Research Center (HEASARC) provided by
NASA's Goddard Space Flight Center.

\end{document}